\documentclass[aps,prl,twocolumn]{revtex4-1}
\usepackage{graphicx}
\usepackage{amsmath}
\usepackage{amssymb}
\usepackage{braket}
\usepackage{hyperref}
\usepackage{subfigure}
\usepackage[usenames,dvipsnames]{color}

\usepackage{graphicx}
\usepackage{subfigure}
\usepackage[mathscr]{euscript}

\newcommand{\ct}{\cite}
\newcommand{\la}{\lambda}
\newcommand{\bi}{\bibitem}
\newcommand{\be}{\begin{equation}}
\newcommand{\ee}{\end{equation}}
\newcommand{\ba}{\begin{eqnarray}}
\newcommand{\ea}{\end{eqnarray}}

\newcommand{\non}{\nonumber}
\newcommand{\Tr}{{\rm Tr}}
\newcommand{\e}{\mathrm{e}}

\begin{document}

\title{Growth of mutual information in a  quenched   one-dimensional open  quantum many-body system}
\author{Somnath Maity, Souvik Bandyopadhyay, Sourav Bhattacharjee and Amit Dutta} 
\affiliation{Department of Physics, Indian Institute of Technology Kanpur, Kanpur 208016, India}

\begin{abstract}
	
We study  the temporal evolution of the  mutual information  (MI) in a one-dimensional Kitaev chain, coupled to a fermionic Markovian bath,  subsequent to a global quench of the
chemical potential. In the unitary case,  the MI (or equivalently the bipartite entanglement entropy) saturates to a steady-state value (obeying a volume law) following a ballistic growth. On the
contrary,  we establish that in the dissipative case the  MI is  exponentially damped both during the initial ballistic growth 
as well as in the approach to the steady state. 
  {We observe that even in a dissipative system, post quench information propagates solely through entangled pairs of  quasi-particles having a finite life time; this quasi-particle picture is further 
corroborated by the out-of-equilibrium analysis of two-point fermionic correlations.  Remarkably, in spite of the finite life time of the quasi-particles,  a finite  steady-state value of the  MI survives  in asymptotic times which  is an artefact of non-vanishing two points correlations.} Further, the finite life time of quasi-particles renders to a finite length scale in these steady-state correlations.
	
\end{abstract}
\maketitle

  The mutual information (MI) is an important measure of the amount of correlations present in a quantum system. For a pure composite state, it is equivalent with the bipartite entanglement entropy (EE), which is the more commonly used tool for probing purely quantum correlations. The EE has a wide range of applications in areas ranging from quantum computation \ct{nielsen10} and quantum many body physics \cite{amico08,latorre09,eisert10} to conformal field theory \cite{calabrese09}, quantum gravity \cite{nishioka09} and black holes \cite{solodukhin11}. In condensed matter physics, for example, the EE can be used to probe quantum criticality \cite{vidal03} and complexity \cite{schuch08}. The scaling behaviour of  the EE enables us to distinguish quantum phases that cannot be characterized by symmetry properties, such as topological phases of matter   \cite{kitaev06,levin06,jiang12} and spin liquids \cite{zhang11,isakov11}. Experimental studies \cite{rajibul15} probing the purity of the reduced system under consideration, firmly establish the entropic measures of many-body entanglement, both in and out of equilibrium, on physical grounds. 
  The underlying aspect that makes the EE relevant in such diverse areas as listed above is intricately tied to the following fundamental question: how do
quantum correlations propagate in a system under different incumbent situations?

If the composite system is pure, the  bipartite EE is calculated through the von Neumman entropy of the reduced density matrix $\rho_\ell$ for a sub-system of
size $\ell$;
\be\label{eq_ee}
S(\ell)=-\Tr\left[\rho_\ell \ln\rho_\ell\right],
\ee
where $\rho_\ell=\Tr_{L-\ell}\left[\rho_L\right]$ and   $\rho_L$ is the density matrix of  the composite system of size $L$. For a short-range one-dimensional Hamiltonian with a gapped spectrum, the  EE follows an area law \cite{vidal03,calabrese04}, while a  logarithmic scaling exists for the gapless phase \ct{vidal03}.

Recently, there has been  great interest in studying the temporal evolution of the EE of an isolated quantum many body system following a quantum quench in which a parameter of the Hamiltonian is changed \cite{calabrese05,polkovnikov11,chiara06,burrell07,eisler07,calabrese07,fagotti08,eisler08,eisler08_1,prosen08,stephan11,divakaran11,igloi12,bardarson12,vosk14,ponte15,rajak16,sen16,sen17,russomanno16,apollaro16,pitsios17,bhattacharya18,maity18,nandy18,cincio07,canovi14}.
The out-of-equilibrium dynamics of the system results in the propagation of quantum correlations  over the whole system thus leading to a growth of  non-trivial bipartite entanglement even if the initial state
is  completely unentangled.
It has  been established   \cite{calabrese05} that  following  a global   quench
 of a one-dimensional free fermionic Hamiltonian,  $S(\ell,t)$ 
exhibits a ballistic growth, $ S(\ell, t) \sim t$,   up to a time $t^*=\ell/2v_{\rm max}$, where   $v_{\rm max}$
is the maximum group velocity of information propagation   {determined by the so-called Lieb-Robinson bound \cite{lieb72}. (see also \ct{comment}). This `light-cone' like spreading of correlations has also been experimentally demonstrated in Ref. \cite{cheneau12}. For $t>t^*$,  $S(\ell, t)$ saturates to a constant value proportional to $\ell$, and hence  the EE satisfies a volume law in the steady-state.
  {

{However, all the studies of the EE carried out so far fundamentally assumes an isolated or non-dissipative setting. This is simply because the EE itself is known to quantify bipartite entanglement when the composite system is pure. On the contrary, the presence of at least minimal dissipation in any physical system is inevitable. The purpose of our work is thus to investigate this unexplored area, namely how do correlations propagate in an out-of-equilibrium quantum many body system in  the presence of a weak {\it dissipative} environment. 
The first necessary step to address in this regard is to define an appropriate measure similar to the EE to probe the propagation of correlations. Since the composite system is in a mixed state in the dissipative situation, the bipartite EE as defined in Eq.~\eqref{eq_ee} is naturally no longer significant as a measure of entanglement unlike 
the unitary case. We therefore define another measure which, as we show, is equivalent to the mutual information (MI) \cite{nielsen10}. Although the MI is not guaranteed to strictly encode only quantum correlations \ct{groisman04}, it nevertheless reproduces the results of the quantum correlations manifested in the  EE at least in the non-dissipative limit, thereby permitting a critical comparison between the two scenarios. Furthermore, the dynamics of MI has also recently been studied in the context of quantum information scrambling after a quantum quench~\cite{alba19}.
The second question we address is whether the dissipative coupling to the bath results in an instantaneous (space like) propagation of correlations, thus violating the Lieb-Robinson bound. Although, for a general Markov process with short-ranged interactions, a Lieb-Robinson-like limit has been shown to exist \ct{poulin10}, and is indeed corroborated by our results, a concrete physical picture, particularly in terms of propagating quasi-particles is not yet established. Finally, we also explore the fate of the correlations in the asymptotic steady-state of the system. In this regard, we note that some works \ct{pizorn08,prosen10} have revealed the existence of short-range as well as long-range correlations in the asymptotic steady states of a dissipative  system; however, the dynamical emergence of these asymptotic steady-state correlations remains unexplored. Our work  addresses these less understood areas which are crucial to gain a better understanding of how the dissipative behavior emerge from the well understood unitary settings.





\textit{Mutual informartion}-- The MI is defined as ~\ct{nielsen10}:
\be
I(\ell:L-\ell)=S\left(\rho_{\ell}\right)+S\left(\rho_{L-\ell}\right)-S\left(\rho_{L}\right),
\label{eq_minfo}
\ee  
where $S\left(\rho_{\ell}\right)$, $S\left(\rho_{L-\ell}\right)$, and $S\left(\rho_{L}\right)$ are the von Neumann  entropies of the sub-system, the rest of the system and the composite  system, respectively. 
For a pure composite system, $S(\rho_L)=0$  and  $S(\rho_{\ell})=S(\rho_{L-\ell})$; consequently, the MI remains equivalent to the bipartite  EE for the unitary evolution, provided the initial state  is pure.  However, for a mixed $\rho_L$,  $S(\rho_{\ell})\neq S(\rho_{L-\ell})\neq I(\ell:L-\ell)/2$.

By splitting the quantity $S\left(\rho_{L}\right)$ into two parts,
we can rewrite Eq.~\eqref{eq_minfo} as
\ba
I(\ell:L-\ell)&=& S'(\ell)+S'(L-\ell),
\ea
where 
\ba\label{eq_eprime}
S'(\ell)&=& S\left(\rho_{\ell}\right)-\frac{\ell}{L}S\left(\rho_{L}\right), \\
S'(L-\ell) &=& S\left(\rho_{L-\ell}\right)-\frac{L-\ell}{L}S\left(\rho_{L}\right).
\ea
Interestingly, in our case,  $S'(\ell) =S'(L-\ell)= I(\ell:L-\ell)/2$  (see  \cite{sm} for verification).
In hindsight, this factorization of $S(\rho_{L})$ is meaningful when  the Lindbladian operators act independently on each site with uniform coupling strengths.


\textit{Model and the bath}-- In this work, we consider a one-dimensional Kitaev chain \textit{globally} coupled to a Markovian bath (see Ref. \cite{sm} for details) and  model the dynamical evolution of this  system within a Lindbladian framework \cite{lindblad76,zoller2000,breuer02}. The Kitaev Hamiltonian~\cite{kitaev01} (see also Ref.~\cite{maity19} and references
therein) we work with is, 
\be
H=-\sum_{n=1}^{L}\left(c_n^\dag c_{n+1} - c_nc_{n+1} + h.c.\right)
-\mu\sum_{n=1}^L\left(2c_n^\dagger c_n-1\right);
\ee
where $c_n$ ($c_n^\dagger$) are the fermionic annihilation (creation) operators acting on the $n$th site. The chain is prepared in the unentangled ground state of the Hamiltonian with
a large negative value of the on-site potential ($\mu$) which is instantaneously quenched to the  final critical value $\mu_f=1$. We note that this model can also be mapped to the transverse field Ising model through a Jordan-Wigner  transformation \ct{lieb61,kogut79, sachdev11,suzuki13,dutta15,comments2}.
Following a Fourier transformation to the quasi-momentum ($k$) basis,
the Hamiltonian decouples to the form: $H=\sum_{k>0}H_k$,
where $H_k$ assumes   a $4 \times 4$ form ({see Ref. \ct{sm}).


We  choose a specific Markovian bath where the system-bath interaction is characterized by  a set of Lindblad operators   $\mathcal{L}_n=c_n, \forall n$ \cite{carmele15,keck17,souvik18,souvik20}. 
 The efficacy of such a choice of Lindbladian operator is that in the case of homogenous coupling strengths ($\kappa_n=\kappa$,  $\forall ~n$),  the Lindblad master equation   (with $\hbar=1$)
gets decoupled in the momentum modes as  
\be
\frac{d\rho_k(t)}{dt}~=~-i\left[H_k,\rho_k(t)\right]+\mathcal{D}_k\left[\rho_k(t)\right], 
\label{lindblad_k}
\ee
with
\ba
\mathcal{D}_k\left[\rho_k(t)\right]&=&\kappa\Big( c_k \rho_k(t)c_k^{\dagger}-\frac{1}{2}\left\{c_k^{\dagger}c_k,\rho_k(t)\right\} \non\\
&&~~~ + c_{-k} \rho_k(t)c_{-k}^{\dagger}-\frac{1}{2}\left\{c_{-k}^{\dagger}c_{-k},\rho_k(t)\right\}\Big),
\label{dissipator_k}
\ea
where  $\rho_k(t)$ is the $4\times 4$ time evolved  density matrix, $c_{k}^\dagger$ is the fermionic creation operator for the mode $k$ and  $\rho(t)=\otimes_{k>0} \rho_k(t), \forall t$.

\textit{Numerical results}--  The system is initially prepared in the unentangled ground  state of the Hamiltonian $H_k(\mu)$  with $\mu\to - \infty$.
 At $t=0+$, the on-site potential $\mu$ is suddenly changed to the critical value $\mu_f=1$ and  the system evolves with the final  Hamiltonian $H(\mu_f)$ following  Eq.~\eqref{lindblad_k}. Numerically evaluating  $\rho_k(t)$, we arrive at the 
two point correlations (TPCs) of fermions;
\ba\label{correlators}
C_{mn}\left(t\right)=\Tr\left[\rho(t),c_m^\dagger c_n\right],~
F_{mn}\left(t\right)=\Tr\left[\rho(t),c_m^{\dagger} c_n^{\dagger}\right],
\ea
where $m,n=1,2,\dots,\ell$. Finally, diagonalizing the $2\ell\times2\ell$ correlation matrix 
\be
\mathbb{C}_\ell~=~\left(\begin{array}{cc} I-C & F\\ F^{\dagger} & C \end{array}\right)
\label{cormat}
\ee  
for the sub-system of size $\ell$,  one can calculate  $S(\rho_{\ell},t)$  and hence the MI ($S'(\ell,t)$) for a free fermionic model~\cite{peschel03} (see Ref. \cite{sm}).

\begin{figure}[t]
	\centering
	\subfigure[]{%
		\includegraphics[width=.45\textwidth,height=5.43cm]{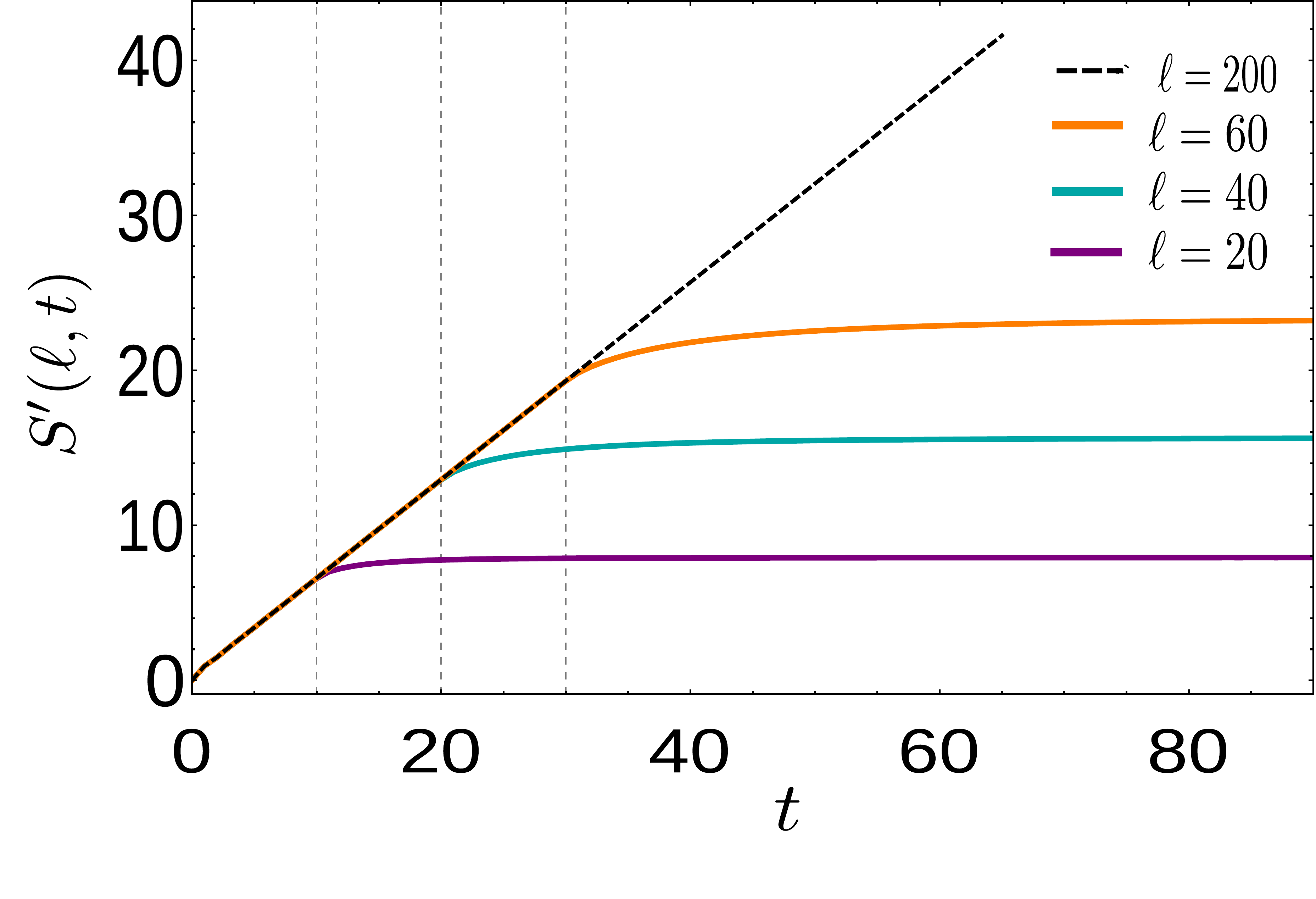}
		\label{fig_close_ee}}
	\hspace{0.1cm}
	\quad
	\subfigure[]{%
		\includegraphics[width=.45\textwidth,height=5.3cm]{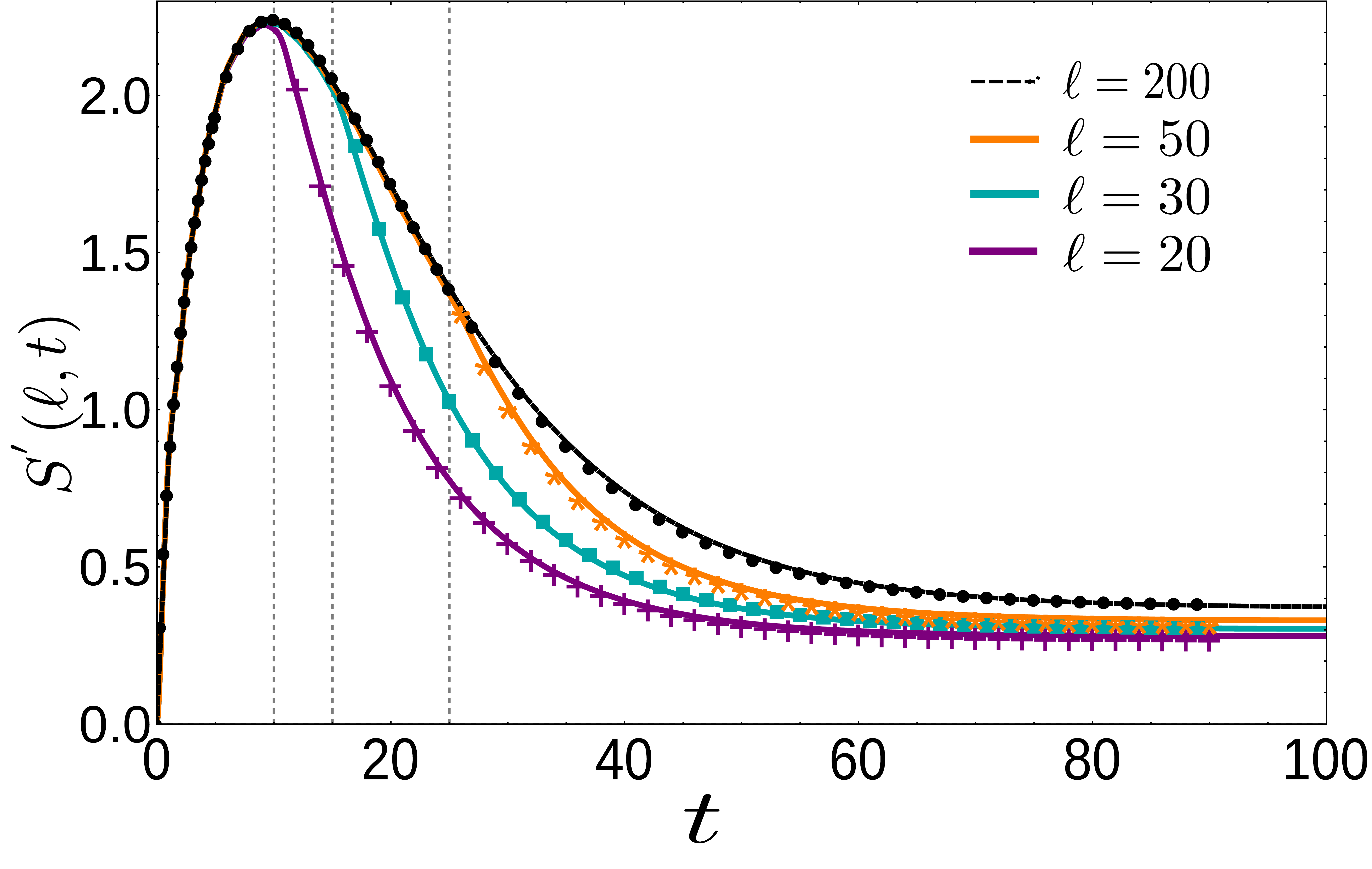}
		\label{fig_open_ee}}
	\caption{ (Color online) The time evolution of $S'(\ell, t)$ for (a) $\kappa=0$ and (b) $\kappa=0.05$ with different sub-system size $\ell$. The black dashed line is for a large $\ell=200$ and indicates  the results in the limit $\ell\rightarrow\infty$ (with $\ell\ll L$). In (b), the markers denote the plots using the fitting functions defined in Eq.~\eqref{eq_large_l} for $t<\ell/2$ and Eq.~\eqref{eq_ee_p} for $t>\ell/2$ as discussed in the text. Here, in both cases (a) and (b), the total system size $L=500$.}
	\label{fig_ee}
\end{figure}

\textit{Unitary situation}:  For  $\kappa=0$ \cite{calabrese05}, the temporal evolution of the MI is  reproduced  in Fig.~\ref{fig_close_ee}.  Indeed, the MI shows a ballistic growth  up to a time  $t^*=\ell/2$ and subsequently saturates to a constant value which is proportional to $\ell$. We note that in the limit $L\rightarrow\infty$, $\ell\rightarrow\infty$   with $\ell\ll L$, the MI  is expected   to have a ballistic growth indefinitely in time. On the contrary,  for any finite $\ell$, the MI follows the $\ell\rightarrow\infty$ line only up to the time $t=t^*=\ell/2$. 
This behaviour of the MI has been  explained by a semiclassical description in terms of entangled pair of  quasi-particles  with equal and opposite  group  velocities $v$ with $v_{\textrm{max}}=1$, resulting in a light-cone
like spread of correlations \cite{lieb72}. 
Further, the steady-state saturation of the MI is an artifact of the infinite life time of the quasi particles, although similar behavior has also been observed in a non-integrable model, where presumably information does not propagate through ballistically moving quasi-particles \cite{kim13}. 

\textit{Dissipative situation:} We now proceed to the case  $\kappa \neq 0$; the temporal variation of  $S'(\ell, t)$, obtained from Eq.~\eqref{eq_eprime} using $\rho_k(t)$,
is presented  in  Fig.~\ref{fig_open_ee} for different sub system sizes $\ell$. Critically inspecting the results,   we conclude that the temporal evolution of $S'(\ell, t)$ is significantly different when compared to that
of the MI   with $\kappa=0$. For a given $\ell$, we observe  a monotonic non-linear growth of the entanglement  after which it eventually  decays to the  steady-state value. The difference with the $\kappa=0$ situation is even more prominent in the limit $L\rightarrow\infty$, $\ell\rightarrow\infty$ with  $\ell\ll L$, where, unlike the indefinite ballistic growth of MI expected  in the former case, $S'(\ell, t)$ is found to
decay to a finite ($\ell$-independent) steady value following the initial growth. However,  similar to the case   $\kappa=0$, the deviation of $S'(\ell, t)$ from  $S'(\ell \to \infty, t)$ again  occurs  at the same instant  $t=\ell/2$.

We propose the the following functional form of $S'(\ell, t)$ in the limit $\ell\rightarrow\infty$,  
\be
S'\left(\ell\rightarrow\infty, t\right)\sim \mathcal{A}\left(1-e^{-c(\kappa)t}\right)+\mathcal{B} t e^{-d(\kappa)t},
\label{eq_large_l}
\ee
with $c(\kappa) \gg d(\kappa)$ and the parameter $\mathcal{B}$ is a non-universal constant depending  on the group velocity of quasi-particles. Eq.~\eqref{eq_large_l} gives a perfect fitting with the numerically obtained results  (see  Fig.~\ref{fig_open_ee}). The functional form can be interpreted in the following way: In early times, $S'\left(\ell\rightarrow\infty, t\right)\sim \mathcal{B}t e^{-d(\kappa)t}$, which essentially means that the linear growth in the MI observed in the $\kappa=0$ situation is now exponentially suppressed. This suppression can be attributed to the finite life-time of the quasi-particles in the case of finite dissipation. However, a finite MI  survives in the asymptotic steady state, $S'(\ell\to\infty,t\to\infty)=\mathcal{A}$.
As   argued below that, surprisingly,  this remanent MI is  
due to perpetual coupling with the bath.

Following the insight obtained from the $\kappa=0$ situation, it is natural to expect that for a finite $\ell$, the early time growth of the MI ($\sim t e^{-d(\kappa)t}$) will continue as long as the quasi-particles originating from the center of the sub-system do not cross the sub-system boundary, following which the MI will only decay exponentially, i.e., $\sim e^{-d(\kappa)t}$. This is what we indeed observe in Fig.~\ref{fig_open_ee}. We therefore arrive at a modified functional form  for  $t >\ell/2$,
\be\label{eq_ee_p}
S'(\ell, t)\sim \mathcal{P}(\ell,\kappa) + \frac{\mathcal{B}\ell}{2} e^{-d(\kappa) t} ~~~~~~~,
\ee
which again fits perfectly with the numerical results [Fig.~\ref{fig_open_ee}].
The  steady state value $\mathcal{P}(\ell, \kappa)$ approaches  $\mathcal{A}$ in the limit $\ell\rightarrow\infty$;  this steady value
satisfies an area law resulting from  a finite correlation length $\xi_k$ defined below.



\textit{Two-point correlations}-- 
 At this point, a  proper analysis of the influence of the bath on the TPCs given in Eq.~\eqref{correlators} is  instrumental in understanding the behavior of the $S'(\ell, t)$ presented
above. We note that the same for a quantum XXZ chain has already been studied \ct{wolf19}.

\begin{figure}[h]
	\centering
	\subfigure[]{%
		\includegraphics[width=.22\textwidth,height=3.2cm]{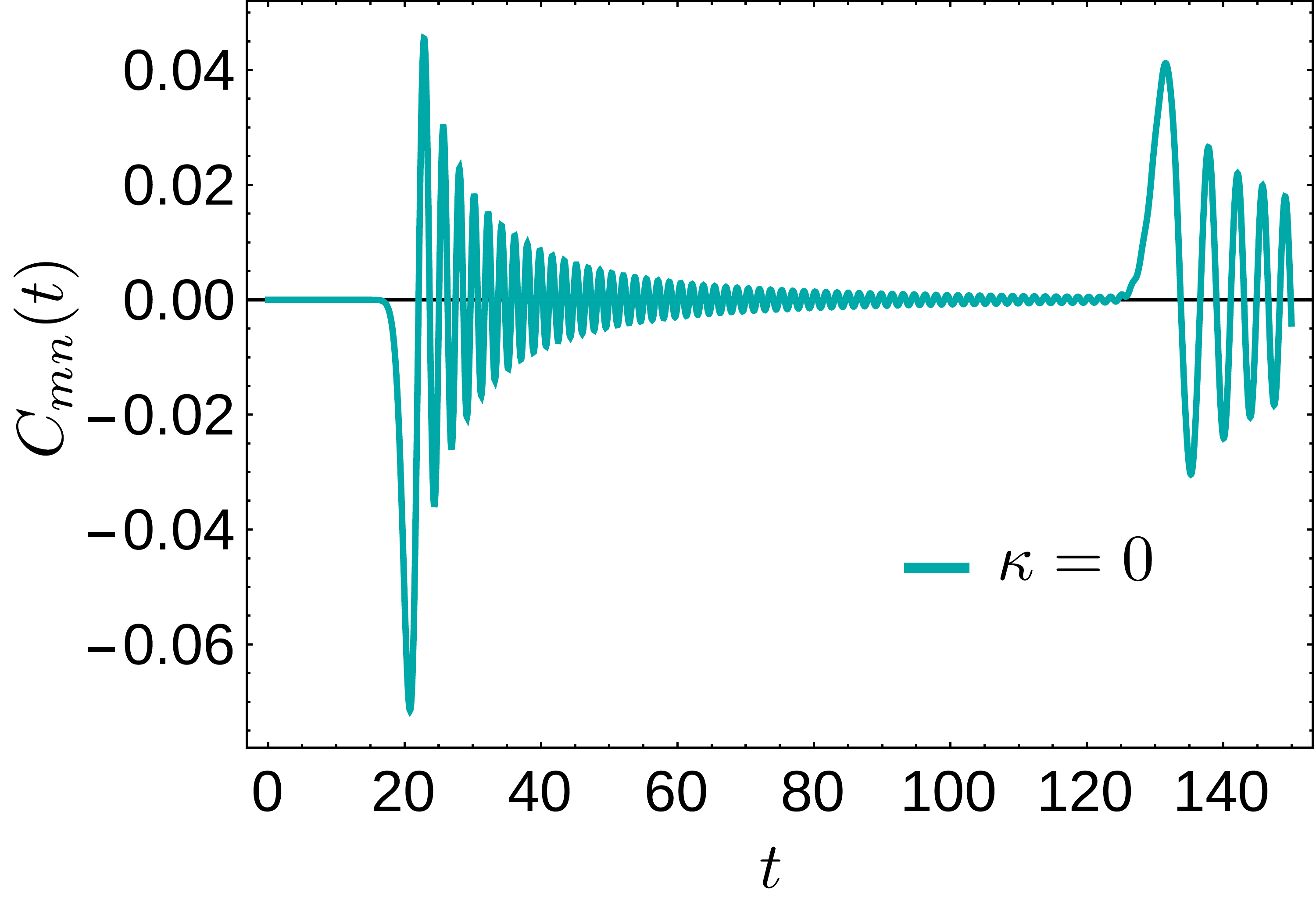}
		\label{fig_tpc_close}}%
	\quad%
	\subfigure[]{%
		\includegraphics[width=.22\textwidth,height=3.2cm]{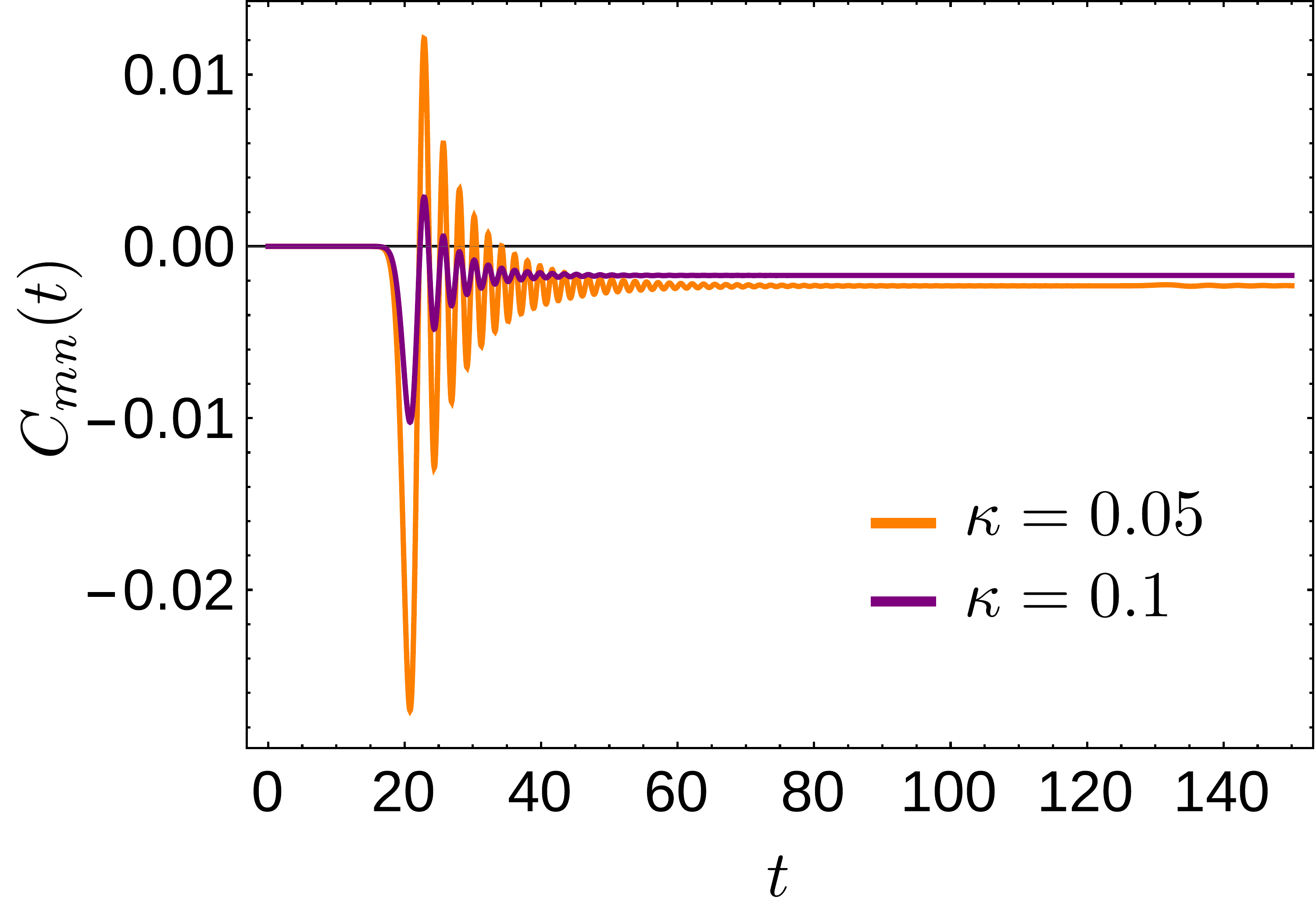}
		\label{fig_tpc_open}}%
	\caption{ (Color online) This figure shows the time evolution of the two point correlation $C_{mn}(t)$ for (a) $\kappa=0$ and (b) $\kappa=0.05, 0.1$ under the quenching protocol ($\mu=-\infty$ to $\mu=1$). We have considered two fixed sites $m$ and $n$ with $|m-n|=40$,  and $L=300$.}
	\label{fig_2ptc}
\end{figure}
For a translationally invariant system with periodic boundary conditions, the single particle TPC, $C_{mn}(t)$, between sites $m$ and $n$, depends on the
distance between the two sites (either  $|m-n|$ or $L-|m-n|$). In Fig.~\ref{fig_2ptc}, we plot the temporal evolution of $C_{mn}(t)$ for different values of $\kappa$. For the unitary case ($\kappa=0$), a finite correlation begins to develop only after a time $t=\textrm{min}\{|(m-n)/2|,|(L-m+n)/2|\}$. 
The existence of a  finite correlation between the two sites `$m$' and `$n$' at a particular instant of time can now be attributed to the arrival  of one quasi-particle of an entangled pair at site $m$ and the other at site $n$ at the same time. This evidently ensures that {\it only} the quasi-particles originating at the middle of either of the two segments between the two points can contribute to TPCs. Since the minimum time taken by these quasi-particles to arrive at the two sites is given by $t^{*}=\textrm{min}\left\{|(m-n)/2|,|(L-m+n)/2|\right\}$, no finite correlation is observed for $t<t^{*}$. The successive peaks occur due to the slower  moving quasi-particles ($v(k)<v_{\textrm{max}}$). The slower moving quasi-particles are less abundant than the faster moving ones, as evident from the progressively decreasing amplitude of the subsequent peaks. A revival of correlation can occur due to the arrival of the quasi-particles from the middle of the longer segment  connecting the two points [see Fig.~\ref{fig_tpc_close}]. We note that in the thermodynamic limit $L\rightarrow\infty$ and for a finite $|m-n|\ll L$, this revival can not occur.

Importantly, even for the case $\kappa\neq0$, no finite correlation is observed for $t<t^*$.  As already mentioned, we note that the maximum velocity of the quasi particles is the same as in the unitary case. The amplitudes of the correlations are also smaller as compared to the $\kappa=0$ case with the amplitude of the revivals significantly diminished. This supports the fact that the quasi particles originating at $t=0$ now have a finite life time. Further, the two-point correlation saturates to a finite non zero value unlike the $\kappa=0$ situation [Fig.~\ref{fig_tpc_open}].  This is manifested also in the non zero steady-state value of the 
$S'(\ell, \infty)$ as analyzed in Ref. \ct{sm}. 

\begin{figure}[t]
	\centering
	\includegraphics[width=.35\textwidth,height=4cm]{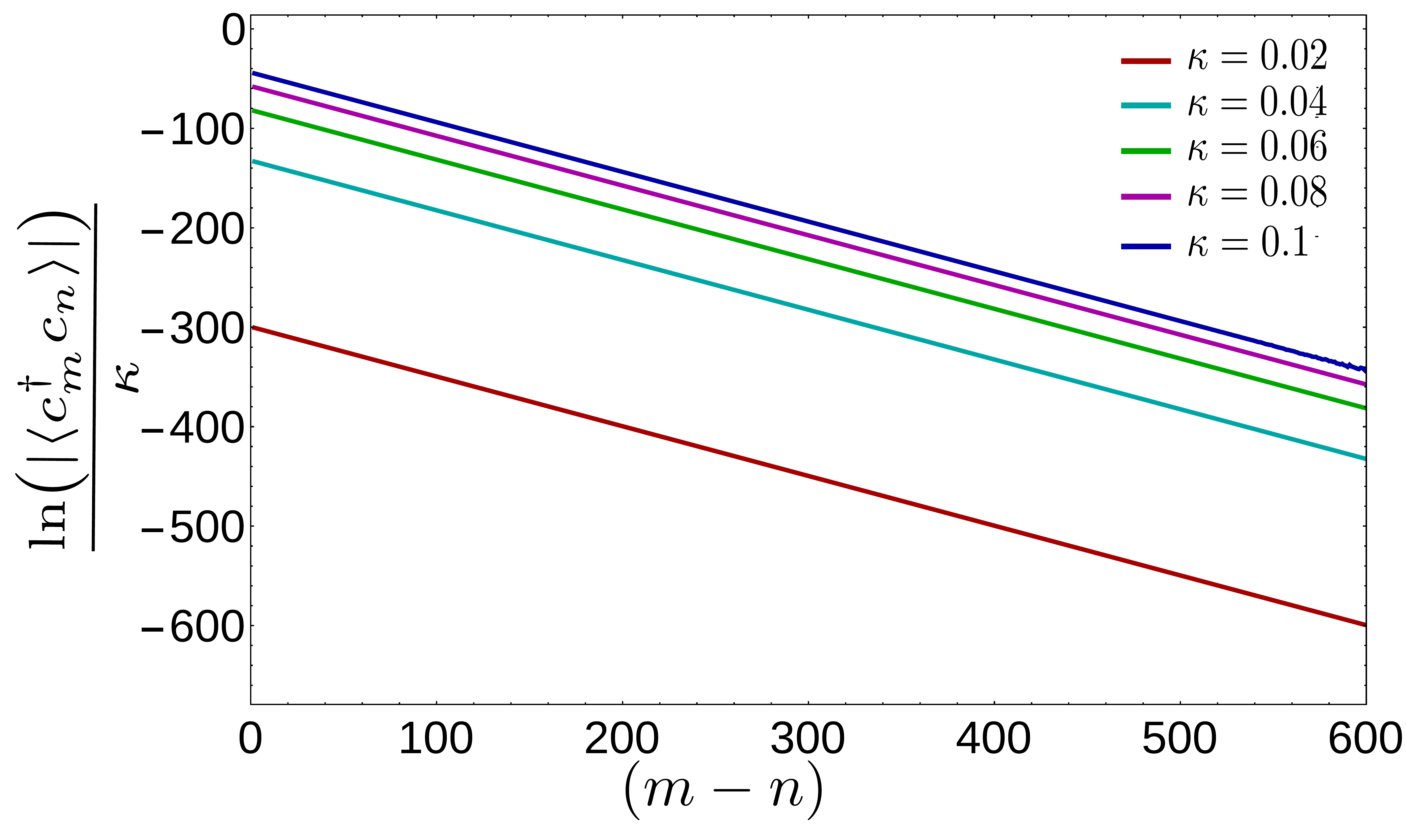}
	\caption{ (Color online) 
	Here,  we have plotted  the quantity $\ln\left( | C_{mn}(\infty) | \right)$ scaled by $\kappa$ with respect to $\left(m-n\right)$. This shows a linear behaviour for all $\kappa$ having a slope   equal to $0.5$,  suggesting  an exponential decay [Eq.~\eqref{eq_sstpc}] of the steady state TPCs.	}
	\label{fig_ss2ptc}
\end{figure}

Finally, the  steady-state
 TPC  when plotted as  a function of the (shorter) intermediate distance $|m-n|$, (see   Fig.~\ref{fig_ss2ptc}) shows  an exponential decay
of the form,
\be\label{eq_sstpc}
C_{mn}(\infty)\sim\exp\left\{-\frac{\kappa(m-n)}{2}\right\}.
\ee
This implies that when  $|n-m|$ exceeds $2/\kappa$, the quasi-particles  generate negligible correlations, thereby pointing to  a finite length scale, $\xi_{\kappa}=2/\kappa$,
 of the steady state TPCs.

The above analysis of the TPCs has the following bearing on the behaviour of $S'(\ell,t)$ discussed earlier. Since the group velocity of the entangled pair of quasi particles remains unaltered for $\kappa \neq 0$,
we observe  that the deviation in $S'(\ell,t)$  from  the $\ell \to \infty$ curve  occurs at $t^* =\ell/2$ as for $\kappa=0$. The remanent TPC in the steady state is reflected in the non zero steady-state value of $S'(\ell,t)$. This steady-state MI , however, results from the collective contribution of remanent TPCs between sites that are located within a finite distance ($\sim \xi_{\kappa})$ from either end of the sub system. The remanent TPCs can be attributed to a \textit{continuous} generation of quasi-particles due to perpetual coupling with the bath as discussed in Ref. \cite{sm}. We note that this is significantly different from the $\kappa=0$ case where the finite MI at large time is maintained collectively by TPCs between all sites of the system.  Further, the finite life time of quasi-particles results in an exponential suppression of the MI at all instants of time.  

 
In summary, we study the influence of a weak dissipative coupling on the generation and growth of the MI in a  quenched one-dimensional integrable model. As in the unitary ($\kappa=0$) case, the MI acquires a finite steady value at long times;  in this case, however, the steady value is determined by both the post quench Hamiltonian as well as the coupling strength with the bath and follows an area law in the thermodynamic limit. Moreover,
the ballistic growth due to the quench-induced propagating correlations is exponentially damped due to the bath-induced dissipation.

To conclude, a scope for further research would be to consider a more general dissipative environment, including thermal as well as non-Markovian baths (see also Ref. \cite{alba20}).
Further, studying the interplay between non-integrability and dissipation
to probe the fate of the ballistic growth in Ref. \ct{kim13} is  a challenging problem. 


\begin{acknowledgments}
 We acknowledge A. Polkovnikov and
D. Sen for fruitful comments. S. Bandyopadhyay acknowledges support from a PMRF fellowship, MHRD, India and
S. Bhattacharjee acknowledges CSIR, India for financial
support. A.D. acknowledges financial support from SPARC
program, MHRD, India.
\end{acknowledgments}

\newpage

\setcounter{equation}{0}
\setcounter{figure}{0}
\setcounter{table}{0}
\setcounter{page}{1}
\makeatletter
\renewcommand{\theequation}{S\arabic{equation}}
\renewcommand{\thefigure}{S\arabic{figure}}
\renewcommand{\bibnumfmt}[1]{[S#1]}
\renewcommand{\@cite}[1]{[S#1]}

\widetext

\begin{center}
	\textbf{\large Supplemental  Material on ``Growth of mutual information in a critically quenched one-dimensional open quantum many body system"} \\
	\vspace{0.5cm}
	\vspace{0.2cm}
\end{center}

\section{Model and the bath} 
In this work, the subsequent  dissipative dynamics of the    Kitaev chain  following a quench of the chemical potential  (from $\mu = -\infty$ and $\mu=1$)  is  assumed to be described by a Lindblad master equation of the form\cite{lindblad76_supp,zoller2000_supp,breuer02_supp},
\be
\frac{d\rho(t)}{dt}~=~-i\left[H,\rho(t)\right]+\mathcal{D}\left[\rho(t)\right];
\label{eq_lindblad_supp}
\ee
we have set $\hbar=1$ througout.
The first term on the right hand side describes the unitary time evolution evolution of the system's density matrix $\rho(t)$ while 
the dissipator $\mathcal{D}\left[\rho(t)\right]$ encapsulates the  dissipative dynamics 
and assumes a form:
\be
\mathcal{D}\left[\rho(t)\right]~=~\sum_n \kappa_n\left(\mathcal{L}_n \rho(t) \mathcal{L}_n^{\dagger} -\frac{1}{2}\left\{\mathcal{L}_n^{\dagger}\mathcal{L}_n,\rho(t)\right\}\right),
\label{dissipator_supp}
\ee 
where $\mathcal{L}_n$ are the Lindblad operators with $\kappa_n$ ($\geq0$)'s  being the corresponding dissipation strengths. 

The   Kitaev chain is described by the Hamiltonian
\be
H=-\sum_{n=1}^{L}\left(c_n^\dag c_{n+1} - c_nc_{n+1} + h.c.\right)
-\mu\sum_{n=1}^L\left(2c_n^\dagger c_n-1\right);
\label{eq_ Kitaev_chain_supp}
\ee 
where $c_n$ ($c_n^{\dagger}$) are Fermionic annihilation (creation) operators residing on the $n$-th site. Further, in the quasi-momentum basis, the Hamiltonian decouples as $H=\sum_{k>0}H_k$, with $k=\left[2\pi(n+1/2)\right]/L$ where $n\in\{0,1, 2 \dots (L-1)\}$. In the basis spanned by the states $\{\ket{\phi^k_1}=\ket{0}, \ket{\phi^k_2}=c_k^\dagger\ket{0}, \ket{\phi^k_3}=c_{-k}^\dagger\ket{0}, \ket{\phi^k_4}=c_k^\dagger c_{-k}^\dagger\ket{0}\}$, $H_k$ assumes the form
\begin{equation}
H_k=\left(
\begin{array}{cccc}
\mu-\cos k & 0 & 0 & \sin k \\
0 & 0 & 0 & 0 \\
0 & 0 & 0 & 0 \\
\sin k & 0 & 0 & -\mu+\cos k \\
\label{ham_k_supp}
\end{array}
\right).
\end{equation}


For a specific system-bath interaction, as used in this work,  characterised by  a set of {\it local} Lindblad operators $\mathcal{L}_n=c_n$ \ct{carmele15_supp,keck17_supp},  the Lindblad master equation given in Eq.~\eqref{eq_lindblad_supp} decouples in the momentum modes as
\be
\frac{d\rho_k(t)}{dt}~=~-i\left[H_k,\rho_k(t)\right]+\mathcal{D}_k\left[\rho_k(t)\right],
\label{lindblad_k_supp}
\ee
with
\ba
\mathcal{D}_k\left[\rho_k(t)\right]=\kappa\left( c_k \rho_k(t)c_k^{\dagger}-\frac{1}{2}\left\{c_k^{\dagger}c_k,\rho_k(t)\right\} + c_{-k} \rho_k(t)c_{-k}^{\dagger}-\frac{1}{2}\left\{c_{-k}^{\dagger}c_{-k},\rho_k(t)\right\}\right).
\label{dissipator_k_supp}
\ea
In the next section, we will cast the Eq.~\eqref{lindblad_k_supp} into a set of coupled differential equations in a particular choice of basis and solve the them numerically to find out $\rho_k(t)$ in the present scenario .

\section{General solution of $\rho(t)$}
In this section, we will briefly elaborate the calculation of the density matrix $\rho_k(t)$ for each momenta mode using the Eq.~\eqref{lindblad_k_supp}\cite{keck17_supp,souvik18_supp}. Let us recall the basis which we considered here, 
\ba\label{eq_basis1_supp}
\ket{\phi^k_1}&\equiv&\ket{0,0},\\
\ket{\phi^k_2}&\equiv&c_k^{\dagger}\ket{0,0}=\ket{k,0},\\
\ket{\phi^k_3}&\equiv&c_{-k}^{\dagger}\ket{0,0}=\ket{0,-k},\\
\ket{\phi^k_4}&\equiv&c_k^{\dagger}c_{-k}^{\dagger}\ket{0,0}=\ket{k,-k},
\label{eq_basis4_supp}
\ea
where $\ket{0,0}$ and $\ket{k,-k}$ refer both the fermionic states for $c_k$ and $c_-k$ are occupied and unoccupied, respectively. In the above basis, we have the following matrix forms of $c_k$, $c_{-k}$ and $H_k(h)$ for each $k$ mode
\begin{equation}
c_k=\left(
\begin{array}{cccc}
0 & 1 & 0 & 0 \\
0 & 0 & 0 & 0 \\
0 & 0 & 0 & -1 \\
0 & 0 & 0 & 0\\
\label{c_k_supp}
\end{array}
\right),
~~~~~~
c_{-k}=\left(
\begin{array}{cccc}
0 & 0 & 1 & 0 \\
0 & 0 & 0 & 1 \\
0 & 0 & 0 & 0 \\
0 & 0 & 0 & 0\\
\end{array}
\right),
~~~\text{and}~~~
H_k(h)=\left(
\begin{array}{cccc}
h-\cos k & 0 & 0 & \sin k \\
0 & 0 & 0 & 0 \\
0 & 0 & 0 & 0 \\
\sin k & 0 & 0 & -h+\cos k \\
\end{array}
\right).
\end{equation}
Using the above equations, the Lindblad equation (Eq.~[10] in the main text) for each $k$ mode  contains sixteen coupled first order linear differential equations of which only ten are independent, those are
\ba
\dot{\rho}_{11}(t)&=&-i\Delta\left(\rho_{14}^{\star}(t)-\rho_{14}(t)\right)+\kappa\left(\rho_{22}(t)+\rho_{33}(t)\right),\non\\
\dot{\rho}_{12}(t)&=& -i\left(\epsilon\rho_{12}(t)+\Delta\rho_{24}^{\star}(t)\right)-\kappa\left(\frac{1}{2}\rho_{12}(t)-\rho_{34}(t)\right),\non\\
\dot{\rho}_{13}(t)&=&-i\left(\epsilon\rho_{13}(t)+\Delta\rho_{34}^{\star}(t)\right)-\kappa\left(\frac{1}{2}\rho_{13}(t)-\rho_{24}(t)\right), \non\\
\dot{\rho}_{14}(t)&=&  -i\left(2\epsilon\rho_{14}(t)+\Delta\rho_{11}(t)+\Delta\rho_{44}(t)\right)-\kappa\rho_{14}(t),\non\\
\dot{\rho}_{22}(t)&=&-\kappa\rho_{22}(t)+\kappa\rho_{44}(t),\non\\
\dot{\rho}_{23}(t)&=&-\kappa\rho_{23}(t),\non \\
\dot{\rho}_{24}(t)&=&-i\left(\epsilon\rho_{24}(t)-\Delta\rho_{12}^{\star}(t)\right)-\frac{3\kappa}{2}\rho_{24}(t),\non\\
\dot{\rho}_{33}(t)&=&-\kappa\rho_{33}(t)+\kappa\rho_{44}(t),\non\\
\dot{\rho}_{34}(t)&=&-i\left(\epsilon\rho_{34}(t)-\Delta\rho_{13}^{\star}(t)\right)-\frac{3\kappa}{2}\rho_{34}(t), \non\\
\dot{\rho}_{44}(t)&=& -i\Delta\left(\rho_{14}(t)-\rho_{14}^{\star}(t)\right)-2\kappa\rho_{44}(t). 
\label{coupled_eqs_supp}
\ea 
where $\epsilon=\mu-\cos k$, $\Delta=\sin k$ and $\dot{\rho}_{ij}=d\rho_{ij}/dt$. Solving the above equations we can obtain $\rho_k(t)$.
It is evident from the above equations that if we consider the initial state  prepared at $\mu_i=-\infty$ i.e., $\rho_k(0)=\ket{0,0}\bra{0,0}$, the density matrix $\rho_k(t)$ has only five non-zero elements, $\rho_{11}(t)$, $\rho_{22}(t)$, $\rho_{33}(t)$, $\rho_{44}(t)$, and $\rho_{14}(t)$. Thus, at any instant of time $t$, the density matrix for each $k$ mode takes the following form
\be\label{app_eq_dmk}
\rho_k(t)=\left(
\begin{array}{cccc}
	\rho_{11}(t) & 0 & 0 & \rho_{14}(t) \\
	0 & \rho_{22}(t) & 0 & 0 \\
	0 & 0 & \rho_{33}(t) & 0 \\
	\rho_{14}^{\star}(t) & 0 & 0 & \rho_{44}(t)\\
\end{array}
\right).
\ee
Having the solutions of $\rho_k(t)$,  we can proceed to calculate the fermionic two point correlations (TPCs) and eventually the mutual information (MI) $S'(\ell,t)$.

\section{Calculation of fermionic two point correlations}\label{app_correlators}
We now present the calculation of fermionic two point  correlation functions using the density matrix $\rho_k(t)$. Let us first write the Fourier transform of fermion $c_n$ for the momenta $k>0$
\ba\label{eq_ft_supp}
c_n=\frac{1}{\sqrt{L}}\sum_{k>0}\left(\e^{-ikn}c_k+\e^{ikn}c_{-k}\right).
\ea 
Using the Eq.~\eqref{eq_ft_supp}, one can write the two point correlations (TPCs) in real space in terms of TPCs in the momentum space as
\ba
\langle c_m^{\dagger}c_n\rangle=\frac{1}{L} \sum_{k_1, k_2 >0} \Big( \e^{i\left(k_1m-k_2n\right)}\langle c_{k_1}^{\dagger}c_{k_2}\rangle + \e^{i\left(k_1m+k_2n\right)}\langle c_{k_1}^{\dagger}c_{-k_2}\rangle+\e^{-i\left(k_1m+k_2n\right)}\langle c_{-k_1}^{\dagger}c_{k_2}\rangle +\e^{-i\left(k_1m-k_2n\right)}\langle c_{-k_1}^{\dagger}c_{-k_2}\rangle\Big), \non\\
\langle c_m^{\dagger}c_n^{\dagger}\rangle=\frac{1}{L} \sum_{k_1, k_2 >0} \Big( \e^{i\left(k_1m+k_2n\right)}\langle c_{k_1}^{\dagger}c_{k_2}^{\dagger}\rangle + \e^{i\left(k_1m-k_2n\right)}\langle c_{k_1}^{\dagger}c_{-k_2}^{\dagger}\rangle+\e^{-i\left(k_1m-k_2n\right)}\langle c_{-k_1}^{\dagger}c_{k_2}^{\dagger}\rangle +\e^{-i\left(k_1m+k_2n\right)}\langle c_{-k_1}^{\dagger}c_{-k_2}^{\dagger}\rangle\Big), 
\label{correlators_real_supp}
\ea
where all the expectation values on the right hand side are taken over the momentum space density matrix $\rho(t)=\otimes_{k>0}\rho_k(t)$ i.e. for a general operator $\hat{O}$;
\be
\langle\hat{O}(t)\rangle~=~\Tr\left[\rho(t) \hat{O}\right]~=~\sum_{k>0}\sum_{i,j=1}^{4} \rho^k_{i,j}(t) \bra{\phi^k_j}\hat{O}\ket{\phi^k_i}.
\label{expectation_app}
\ee
Using the Eq.~\eqref{expectation_app} and Eqs.~\eqref{eq_basis1_supp}-\eqref{eq_basis4_supp}, one can calculate the following correlations in momentum space;
\ba\label{correlators1_supp}
\langle c_{k_1}^{\dagger}c_{k_2}\rangle&=&\sum_{k>0}\sum_{i,j=1}^{4} \rho^k_{i,j}\bra{\phi^k_j}c_{k_1}^{\dagger}c_{k_2}\ket{\phi^k_i}
~=~\sum_{k>0} \left(\rho^k_{22}+\rho^k_{44}\right)\delta_{k_1,k_2}\delta_{k_1,k}, \\
\langle c_{-k_1}^{\dagger}c_{-k_2}\rangle&=&\sum_{k>0} \left(\rho^k_{33}+\rho^k_{44}\right)\delta_{k_1,k_2}\delta_{k_1,k},\\
\langle c_{k_1}^{\dagger}c_{-k_2}\rangle&=&\sum_{k>0} \left(\rho^k_{32}\right)\delta_{k_2,k}\delta_{k_1,k},\\
\langle c_{-k_1}^{\dagger}c_{k_2}\rangle&=&\sum_{k>0} \left(\rho^k_{23}\right)\delta_{k_2,k}\delta_{k_1,k},
\ea  
and
\ba
\langle c_{k_1}^{\dagger}c_{k_2}^{\dagger}\rangle&=&0,\\
\langle c_{-k_1}^{\dagger}c_{-k_2}^{\dagger}\rangle&=&0,\\
\langle c_{k_1}^{\dagger}c_{-k_2}^{\dagger}\rangle&=&\sum_{k>0} \left(\rho^k_{14}\right)\delta_{k_2,k}\delta_{k_1,k},\\
\langle c_{-k_1}^{\dagger}c_{k_2}^{\dagger}\rangle&=&-\sum_{k>0} \left(\rho^k_{14}\right)\delta_{k_2,k}\delta_{k_1,k}.
\label{correlators2_supp}
\ea
Note that in the last equation the negative sign comes from fermionic anti-commutation relations i.e. using the fact $c_{-k}^{\dagger}c_{k}^{\dagger}\ket{0,0}=-\ket{k,-k}$. Finally, using the above correlation functions in Eqs.~\eqref{correlators1_supp}-\eqref{correlators2_supp} and Eq.~\eqref{correlators_real_supp}, the elements in the correlation matrix $\mathcal{C}$ for the sub system of size $\ell$ can be found as
\ba\label{eq_tpcs_supp_c}
C_{mn}(t)&=&\frac{1}{L}\sum_{k>0} \label{app_eq_2ptc} \left(\rho^k_{22}+\rho^k_{44}\right)\e^{ik\left(m-n\right)} + \rho^k_{32}\e^{ik\left(m+n\right)} + \rho^k_{23}\e^{-ik\left(m+n\right)} +\left(\rho^k_{33}+\rho^k_{44}\right)\e^{-ik\left(m-n\right)} \non \\ 
&=&\frac{2}{L}\sum_{k>0}\rho^k_{44} \cos\left(k(m-n)\right) + \frac{1}{L} \sum_{k>0} \left\{\rho^k_{22}\e^{ik\left(m-n\right)} + \rho^k_{33}\e^{-ik\left(m-n\right)}\right\}\\
F_{mn}(t)&=&\frac{2}{L}\sum_{k>0} \rho^k_{14} \sin\left(k(m-n)\right) \label{eq_tpcs_supp_f}
\ea
where $m,n=1,2,3,\dots\ell$ and the time dependence comes from the time dependent matrix elements $\rho^k_{ij}(t)$ of the density matrix $\rho_k(t)$. Note that in arriving at the final
form of the  Eq.~\eqref{app_eq_2ptc}, we have used the fact that $\rho^k_{23}(t)=\rho^k_{32}(t)=0$ valid for the present choice of initial conditions (see Eq.~\eqref{app_eq_dmk}).  In the unitary case ($\kappa=0$),  $\rho_{22}^{k}(t)$ and $\rho_{33}^{k}(t)$ vanish, rendering a simpler form of the $C_{mn}(t)$.

In our numerical scheme, by solving Eqs.~\eqref{coupled_eqs_supp} ,we shall construct the density matrix given in the Eq.~\eqref{app_eq_dmk} and hence evaluate Eq.~\eqref{eq_tpcs_supp_c} and Eq.~\eqref{eq_tpcs_supp_f}. Having the values of TPCs $C_{mn}(t)$ and $F_{mn}(t)$, we construct the correlation (or covariance) matrix $\mathbb{C}_\ell$ (Eq.~[12] in the main text) and proceed to calculate the von-Neumann entropy $S(\rho_l)$ of the reduced density matrix $\rho_{\ell}$ of a sub-system of size $\ell$ as
\be
S(\rho_{\ell})=-\sum_{i=1}^{2\ell}\la_i\ln\la_i
\ee
where $\la_i$ are the eigenvalues of the correlation matrix $\mathbb{C}_\ell$~\cite{peschel03_supp}.
\begin{figure}[]
	\centering
	\includegraphics[width=.45\textwidth,height=5.cm]{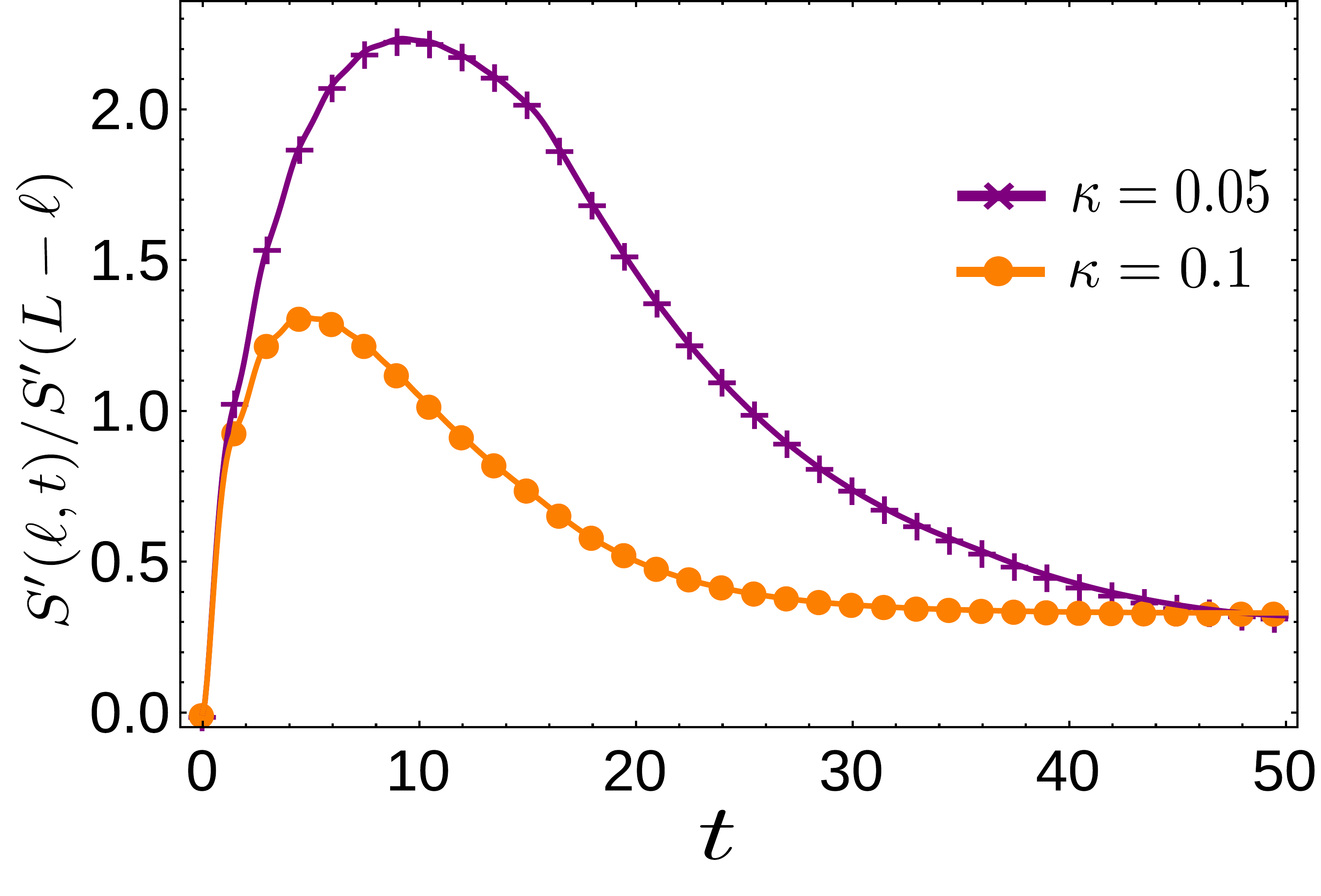}
	\caption{ (Color online) This figure shows that $S'(\ell,t)$ (denoted by solid lines) is equal to $S'(L-\ell,t)$ (denoted by markers) for all instants of time. We have chosen two values of the coupling strength $\kappa=0.05$ and $0.1$,  the  sub-system size $\ell=30$ and the composite system size $L=100$}
	\label{fig_comparison}
\end{figure}
\section{Numerical verification of $S'(\ell, t)=S'(L-\ell, t)$ for all $t$}
Let us recall, the definition of the mutual information (MI)
\be
I(\ell:L-\ell)=S\left(\rho_{\ell}\right)+S\left(\rho_{L-\ell}\right)-S\left(\rho_{L}\right),
\label{eq_minfo_supp}
\ee  
where $S\left(\rho_{\ell}\right)$, $S\left(\rho_{L-\ell}\right)$ and $S\left(\rho_{L}\right)$ are the von-Neumann  entropy of the sub-system, rest of the system and the composite  system, respectively. By splitting the quantity $S\left(\rho_{L}\right)$ into two parts,
we can rewrite the above equation in following way
\ba
I(\ell:L-\ell)&=& S'(\ell)+S'(L-\ell)
\ea
where 
\ba
S'(\ell)&=& S\left(\rho_{\ell}\right)-\frac{\ell}{L}S\left(\rho_{L}\right), ~~~~\text{and}~~~~
S'(L-\ell) = S\left(\rho_{L-\ell}\right)-\frac{L-\ell}{L}S\left(\rho_{L}\right).
\label{eq_eprime_supp}
\ea
Interestingly, for the bath chosen in the present work, the two quantity $S'(\ell)$ and $S'(L-\ell)$ are exactly equal in all time as shown in the Fig.~\ref{fig_comparison}.

\section{Steady state solution of $\rho_k(t)$}
In the presence of the dissipative environment, following a sudden quench the system reaches a steady state in the asymptotic limit of time $t\rightarrow\infty$. In the present scenario, analytical form of the steady state density matrix $\rho_k(\infty)$ for each $k$ mode can be exactly calculated by putting $d\rho_k(t)/dt=0$ in the Lindblad master equation~\eqref{lindblad_k_supp}. The non-zero elements of the density matrix $\rho_k(\infty)$ assume the following  form:
\ba
\rho_{11}(\infty)&=&\frac{\Delta^2+4\epsilon^2+\kappa^2}{4\left(\Delta^2+\epsilon^2\right)+\kappa^2}, \\
\rho_{22}(\infty)&=&\rho_{33}(\infty)=\rho_{44}(\infty)=\frac{\Delta^2}{4\left(\Delta^2+\epsilon^2\right)+\kappa^2},\\
\rho_{14}(\infty)&=&\frac{\Delta\left(2\epsilon-i\kappa\right)}{4\left(\Delta^2+\epsilon^2\right)+\kappa^2},
\ea
where $\epsilon=\mu_f-\cos k$ and $\Delta=\sin k$.  These solutions enable us  to  calculate the steady state two point correlations  $C_{mn}(\infty)$, $F_{mn}(\infty)$ and  the steady state von Neumann entropy $S(\ell,\infty)$ of a sub-system of size $\ell$. Further, the steady state MI $S'(\ell,t)$, calculated using Eq.~\eqref{eq_eprime_supp}, is shown in the Fig.~\ref{fig_ssee_supp}; this steady state value satisfies
	an area law in the thermodynamic limit.
\begin{figure}[]
	\centering
	\includegraphics[width=.45\textwidth,height=5.cm]{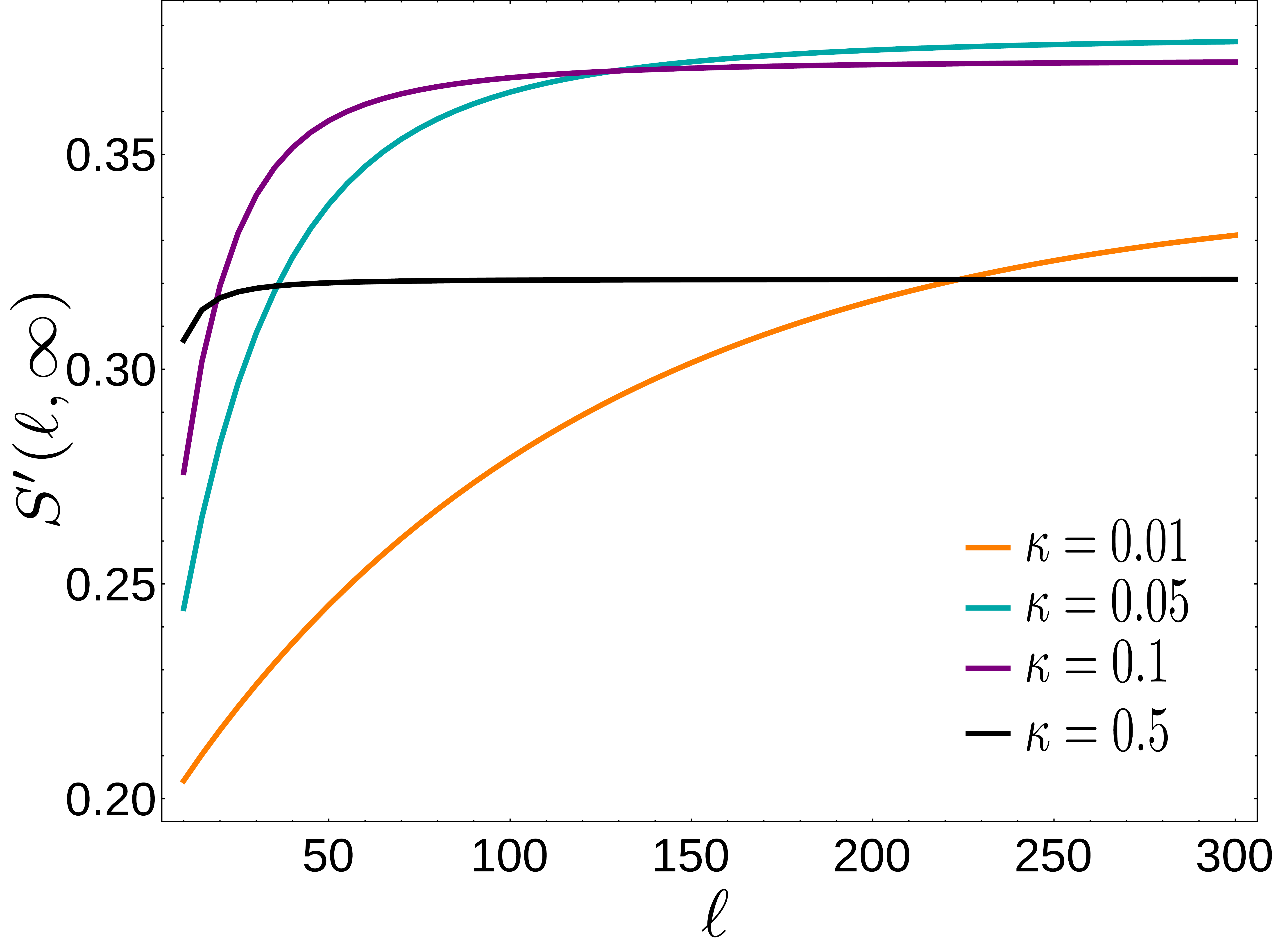}
	\caption{ (Color online) The variation of the steady state value of MI $S'(\ell,t)$  (i.e., $\mathcal{P} (\ell, \kappa)$ in Eq.~[12] of the main text ) with  $\ell$ for different values of $\kappa$. 
		As
		$\kappa$ increases from $0$ to higher values, $\mathcal{P}(\ell,\kappa)$ apparently shows a sub-volume behavior.  In the thermodynamic
			limit the quantity however satisfies an area law as a consequence of finite $\xi_k$..
		Here, we choose $L=1000$.}
	\label{fig_ssee_supp}
\end{figure}

\begin{figure}[]
	\centering
	\subfigure[]{%
		\includegraphics[width=.45\textwidth,height=5.43cm]{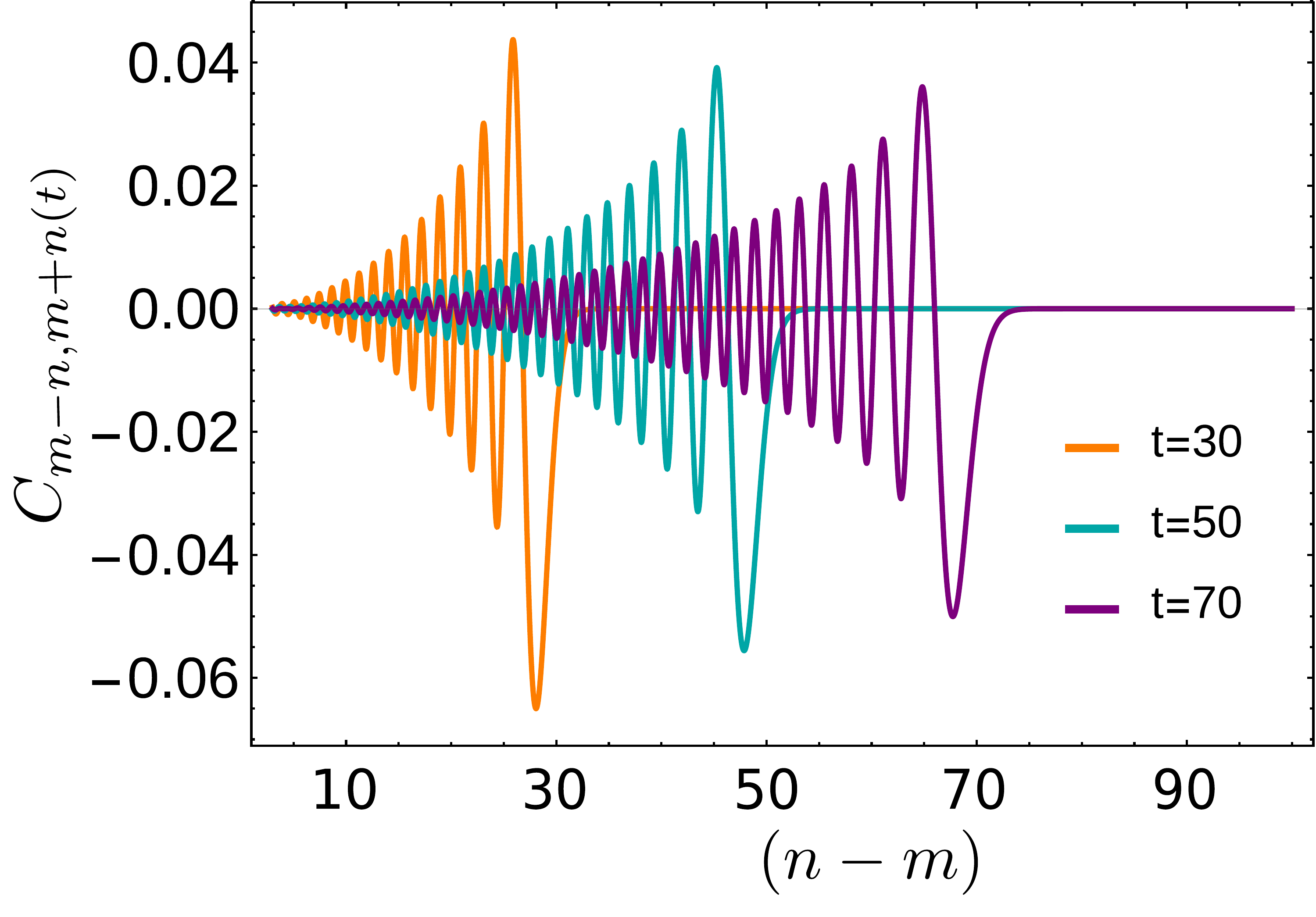}
		\label{fig_app_pairwise_close}}
	\hspace{0.1cm}
	\quad
	\subfigure[]{%
		\includegraphics[width=.47\textwidth,height=5.3cm]{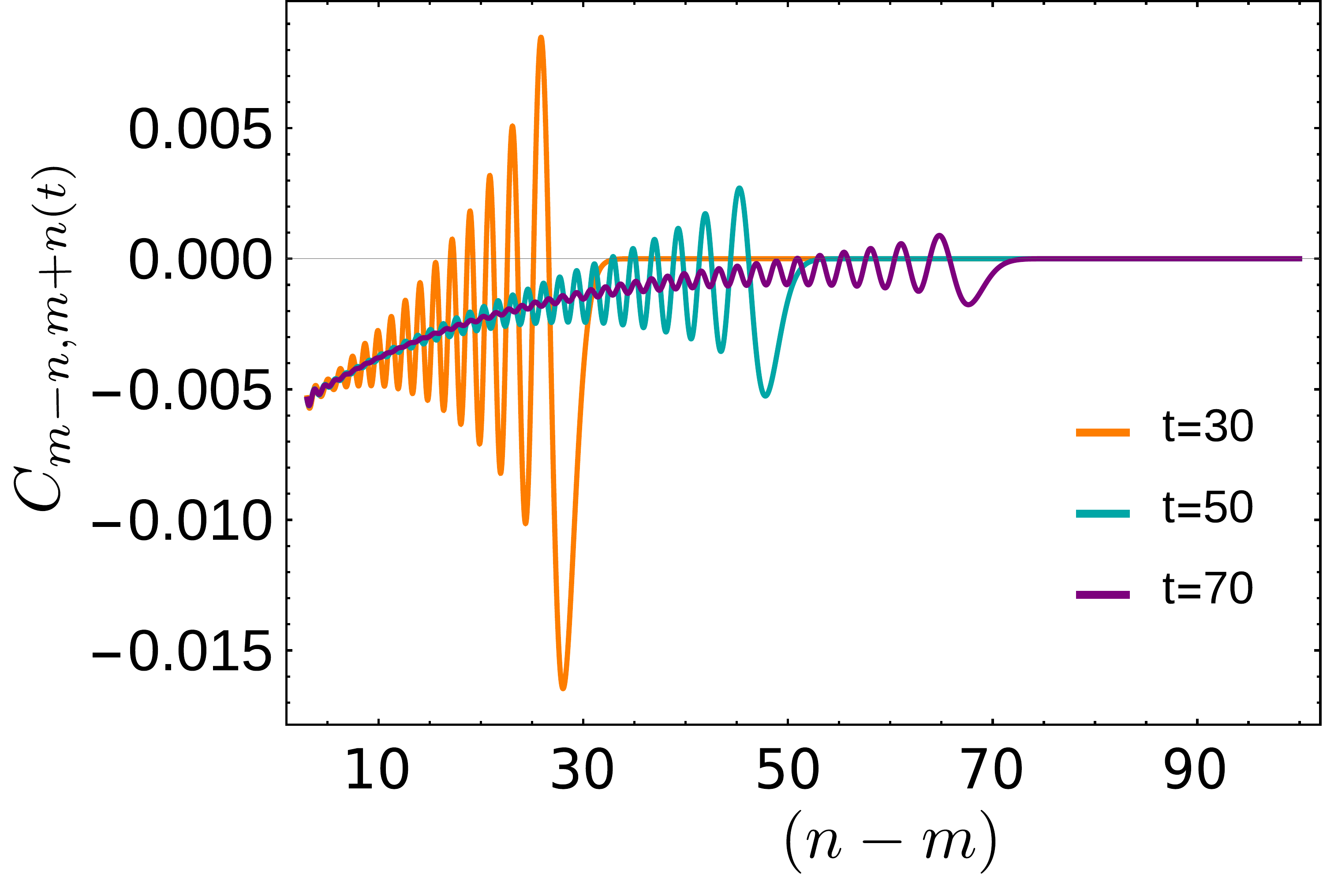}
		\label{fig_app pairwise_open}}
	\caption{ (Color online). Here, we plot the two-point correlations $C_{mn}(t)$ between pairwise points (labelled by $n$)  equidistant from the site $m$ (chosen to be $m=50$) for different instants of time after the sudden quench of the chemical potential
		for  (a) the unitary case ($\kappa=0$) and (b) the dissipative case ($\kappa\neq0$).
	}
	\label{fig_app_pairwise}
\end{figure}
\section{Steady state MI   from two-point correlations}
In this section, we study the dynamics of two-point correlations (TPCs) $C_{mn}(t)$ (given in the Eq.~\eqref{eq_tpcs_supp_c}) further to understand the origin of finite non-zero value of the steady state MI for the dissipative case. Here, we calculate the equal-time single particle  TPCs between pairwise points equidistant from  a fixed site  (denoted by  $m$);
\be
C_{m-n,m+n}(t)=\langle c_{m-n}^{\dagger}(t) c_{m+n}(t)\rangle
\ee
by varying $n$ ($\geq 1$). In Fig.~\ref{fig_app_pairwise}, we plot $C_{m-n,m+n}(t)$ as a function of $(n-m)$ for different instants of time. The two points $m+n$ and $m-n$ become correlated only when entangled pair of quasi-particles reach the two points simultaneously. This essentially implies that only the quasi-particles originating at the mid point $m$ can contribute to $C_{m-n,m+n}(t)$ at any time $t>0$. Note that we have ignored  possible contributions from the mid-point of the other segment $(L-2n)$ due to the circular geometry of the chain; this is a valid approximation as  $L\gg2n$. Thus, it is clear that, at a given time $t$, only the points lying within distance $m-n= v_g t= t$ can have finite pair-wise correlation with their corresponding counter parts in the segment $m+n$, as is clearly seen in Fig~\ref{fig_app_pairwise}. As discussed in the main text, the subsequent peaks after the first peak, i.e. within the region $|n-m|<t$, is due to the presence of the slower moving quasi-particles. In the unitary case (see Fig.~\ref{fig_app_pairwise_close}), the resultant correlation profile evolves in a packet like fashion with its front end propagating with maximum group velocity ($v_{max}=1$) while the tail end elongates in time due to the velocity differences between quasi-particles. In the dissipative case, however, the correlation profile evolves with a non-vanishing tail even though its front end continue to propagate with velocity $v_{max}$ but with a rapidly diminishing amplitude (see Fig. S3(b)). It is important to realise that the finite MI observed in the steady state (see Fig. [1] of the main text) is an artefact of this non-vanishing tail of the correlation profile. Further, the surviving TPCs in the steady state is an exponentially decaying function (see Fig. [3] in the main text) with the intermediate distance $|n-m|$ between the two points suggesting the existence of a finite correlation length, $\xi_k=2/\kappa$, beyond which the TPCs in the steady state have negligible contributions. We have also checked that the real and imaginary part of the other TPC $F_{mn}(t)$ given in Eq.~\eqref{eq_tpcs_supp_f} also has the same dynamical behavior as discussed  above.

The scaling of the MI in the steady state with the sub-system size $\ell$ (see Fig.~\ref{fig_ssee_supp}) is now explained as follows.
	In the steady state, only the points inside the sub-system which are located within a distance $\xi_\kappa$ from the boundary of the sub-system become correlated with the rest of the system.  The two relevant length scales in the steady state are therefore $\xi_\kappa$ and $\ell$. If  $\ell\ll\xi_\kappa$, all the points inside the sub-system, contribute to the steady state MI; consequently it  {\it apparently} follows a volume law in this limit. On the other limit, if $\ell\gg\xi_\kappa$, only the points located in the vicinity of the boundary of the sub-system contribute to the steady state MI thus leading to an area law ($\ell$ independent) behavior in the thermodynamic limit. In the intermediate case when $\ell$ is comparable with $\xi_\kappa$, the steady state MI shows a sub-volume behaviour, as can be seen in Fig.~\ref{fig_ssee_supp}. 

\begin{figure}[]
	\centering
	\includegraphics[width=.45\textwidth,height=5.cm]{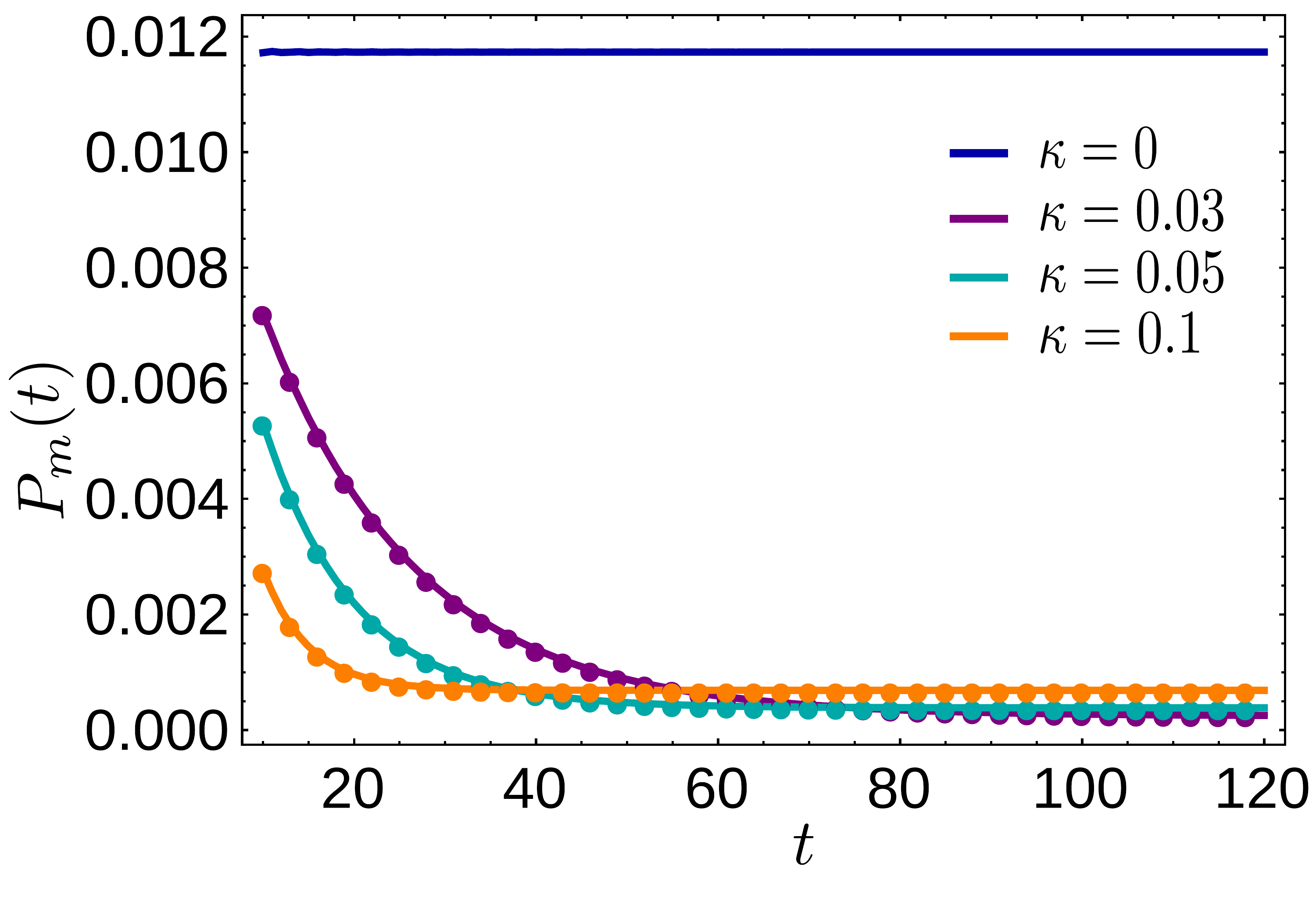}
	\caption{ (Color online) Here, we plot the quantity $P_{m}(t)$ with time $t$ for different values of $\kappa$.  The markers on curves for $\kappa\neq0$ are obtained using the fitted function $\alpha(\kappa)+\beta(\kappa)\exp(-2\kappa t)$. Here, we have chosen the total system size $L=1000$.}
	\label{fig_app_qptotal}
\end{figure}

Finally, we note that the existence of a non-vanishing value of the MI in the steady state despite the quasi-particles having a finite life-time is a striking result in presence of dissipation. We suspect that this happens because the action of the bath is similar to a \textit{continuous} quench on the system which results in generation of quasi-particles at all times. In the steady state, a balance is reached between the generation and destruction processes of the quasi-particles which result in the steady non-vanishing value. Although a rigorous analytical proof of this claim is not feasible, we however, adopt an indirect way to support our claim as follows.

We recall (in the Fig.~\ref{fig_app_pairwise}) that as the system evolves, the amplitude of  oscillations of TPCs decreases rapidly in the dissipative case unlike the unitary case. To capture this difference in a more compact way, we calculate the following quantity 
	\be
	P_m(t)=\sum_{n=m}^{m+v_{max}t} \lvert\langle c_{m-n}^{\dagger}(t) c_{m+n}(t)\rangle\rvert^2,\hspace{0.5cm} v_{max}=1,
	\ee
which at a given time  
can be considered to be a measure of the total number of quasi-particles originated from the point $m$. In the Fig.~\ref{fig_app_qptotal}, we follow the temporal evolution of $P_m(t)$ for different values of $\kappa$. In the unitary case ($\kappa=0$), it remains constant with time which indicates that total number of quasi-particles remains constant for all time; this further implies that the quasi-particles originated by the sudden quench of the chemical potential have infinite life time. However, in the dissipative case ($\kappa\neq0$), $P_{m}(t)$ shows an exponential decay to a constant non-zero steady state value. This exponential decay can be interpreted as the result of the decay of the quasi-particles generated due to the sudden quench at $t=0$. Therefore, for $\kappa\neq0$, a time scale $\tau_\kappa=1/2\kappa$ (other than $t^*$) emerges after which the generation of quasi-particles due to the persistent coupling to the bath and their destruction balances each other, which manifests in the form of a steady value for the MI. 

%
\begin{figure}[]
	\centering
	\includegraphics[width=.47\textwidth,height=5.3cm]{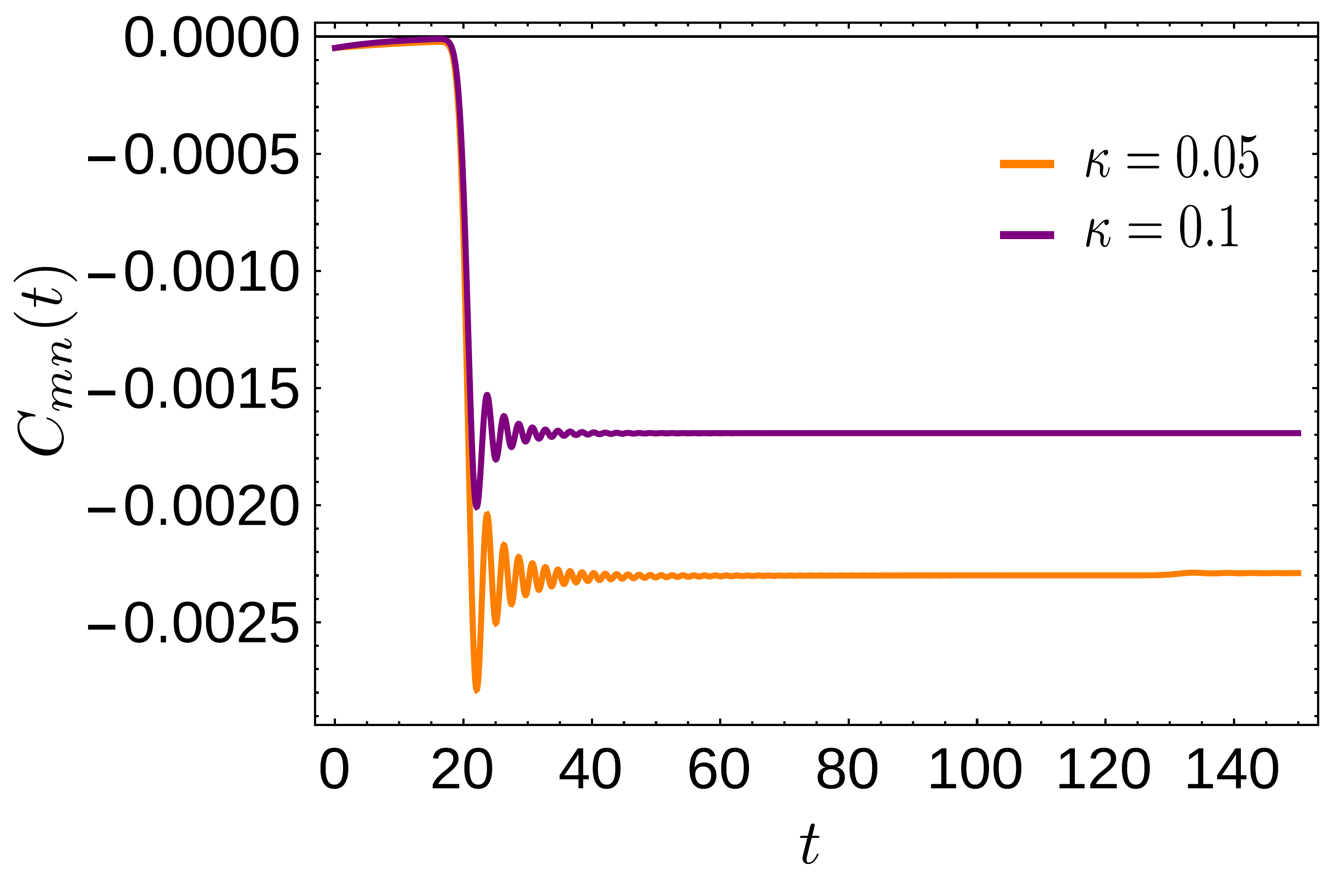}
	\label{fig_2ptc_no_quench}
	\caption{ This figure shows the time evolution of the two point correlation $C_{mn}(t)$ for different values of $\kappa$ under no quench situation for different values of $(m-n)$.  This figure
		should be compared with Fig.~[3] of the main text.}
	\label{fig_no_quench_supp}
\end{figure}

\section{Role of bath}\label{bath_quenching}

To comprehend the role played by the bath, we investigate  $C_{mn}(t)$ with $m$ and $n$ fixed as a function of time  in the
{\it no-quench} situation. In the no-quench situation, there is  no quenching of $\mu$  and 
the   Kitaev chain is initially prepared in the ground state of Hamiltonian $H_k(\mu)$ with field $\mu=1$, which thereafter evolves with  the same Hamiltonian. 
After the decay of the  (small) initial correlation of the critical ground state, we once again observe that a finite correlation starts developing for $t>t^*=\ell/2$ as depicted in Fig.~\ref{fig_no_quench_supp}. This, in hindsight, suggests that the action of attaching the bath is similar to a global quench that results in the generation of quasi-particles throughout the system.

\end{document}